\RequirePackage{fix-cm}
\documentclass[smallextended]{svjour3}       
\smartqed  
\usepackage{graphicx}
 \usepackage{mathptmx}      
\usepackage{amsmath}
\journalname{General Relativity and Gravitation}
\begin{document}

\title{Neutron stars in Scalar-Tensor-Vector Gravity\thanks{This work was supported by grants PICT 2012-00878 (Agencia Nacional de Promoci\'on Cient\'ifica y Tecnol\'ogica, Argentina) and AYA 2013-47447-C3-1-P (Ministro de Educaci\'on, Cultura y Deporte, Espa\~na). We would like to thank Federico Garc\'ia and Santiago del Palacio for helpful comments.}
}

\titlerunning{Neutron Stars in STVG}        

\author{Federico G. Lopez Armengol         \and
        Gustavo E. Romero 
}


\institute{Federico G. Lopez Armengol \and Gustavo E. Romero \at
              Instituto Argentino de Radioastronom\'ia CCT La Plata (CONICET), C.C.5 1894 Villa Elisa, Buenos Aires, Argentina. Tel.: (+54) (0221) 425 4909, Fax:  (+54) (0221) 425 4909 \\
              \email{flopezar@iar-conicet.gov.ar, romero@iar-conicet.gov.ar}
           \and
           Gustavo E. Romero \at
              Facultad de Ciencias Astron\'omicas y Geof\'isicas, Universidad Nacional de La Plata, Paseo del Bosque s/n, 1900 La Plata, Buenos Aires, Argentina. Tel.:  (+54) (0221)-423-6593, Fax: (+54) (0221)-423-6591
}

\date{Received: date / Accepted: date}

\maketitle

\begin{abstract}
Scalar-Tensor-Vector Gravity (STVG), also referred as MOdified Gravity (MOG), is an alternative theory of the gravitational interaction. Its weak field approximation has been successfully used to described Solar System observations, galaxy rotation curves, dynamics of clusters of galaxies, and cosmological data, without the imposition of dark components. The theory was formulated by John Moffat in 2006. In this work, we derive matter-sourced solutions of STVG and construct neutron star models. We aim at exploring STVG predictions about stellar structure in the strong gravity regime. Specifically, we represent spacetime with a static, spherically symmetric manifold, and model the stellar matter content with a perfect fluid energy-momentum tensor. We then derive the modified Tolman-Oppenheimer-Volkoff equation in STVG and integrate it for different equations of state. We find that STVG allows heavier neutron stars than General Relativity (GR). Maximum masses depend on a normalized parameter that quantifies the deviation from GR. The theory exhibits unusual predictions for extreme values of this parameter. We conclude that STVG admits suitable spherically symmetric solutions with matter sources, relevant for stellar structure. Since recent determinations of neutron stars masses violate some GR predictions, STVG appears as a viable candidate for a new gravity theory.
\end{abstract}

\section{Introduction}
\label{s_intro}
	The paramount current theory of the gravitational interaction is General Relativity (GR). The theory earned this status because of its conceptual simplicity, its intuitive geometrical interpretation and, of course, its remarkable successes in modeling gravitational phenomena. But, above the sympathy that these attributes may gender, we must accept that the theory is defective. On a philosophical ground, GR is imperfect as any mathematical representation of reality \cite{bunge1977}. Regarding its physical predictions, GR fails in reproducing rotation curves of galaxies, mass profiles of galaxy clusters, gravitational lensing effects, cosmological data, and spacetime singularities. Some of these problems can be solved or circumvented assuming the existence of dark matter. However, every experiment aimed at measuring the properties of this kind of matter has failed in its quest \cite{aprile2012,lux2013,agnese2014}. In this scenario, alternative gravity theories that do not require the existence of dark components deserve study. \par
 
	In this regard, M. Milgrom \cite{milgrom1983} proposed the MOdified Newtonian Dynamics theory (MOND) to account for astrophysical phenomena without dark matter. MOND correctly reproduces galactic scale observations, but the theory is deficient at galaxy cluster and cosmological scales. Relativistic theories whose weak field limit coincides with MOND were soon formulated. These theories were characterized by the insertion of scalar fields that couple with matter. We highlight the contributions of J. Bekenstein \cite{bekenstein1984} and R. Sanders \cite{sanders1988}. Afterwards, Sanders noticed that any Scalar-Tensor theory fails in explaining light deflection observations without dark matter, so he proposed the incorporation of vector fields to solve the problem \cite{sanders1997}. His ideas matured with the Tensorial-Vectorial-Scalar (TeVeS) theory of gravitation developed by Bekenstein \cite{bekenstein2004}. TeVeS is an admissible relativistic generalization of MOND and a valuable alternative gravity theory. For a review of MOND predictions and its relativistic extensions, see Ref. \cite{famaey2012}.\par
	
	Independently, J. Moffat \cite{moffat2006} postulated the Scalar-Tensor-Vector Gravity theory (STVG), also referred as MOdified Gravity theory (MOG) in the literature. In STVG, the gravitational coupling constant $G$ is reified to a scalar field whose numerical value usually exceeds Newton's constant $G_{\mathrm{N}}$. This assumption serves to describe correctly galaxy rotation curves \cite{brownstein2006}, clusters dynamics \cite{moffat2014}, Bullet Cluster phenomena \cite{brownstein2007}, and cosmological data \cite{moffat2007}, without requiring the existence of dark contributions. In order to counteract the enhanced gravitational coupling constant on Solar System scales, Moffat proposed a gravitational repulsive Yukawa vector field $\phi^{\mu}$. In this way, Newton's gravitational constant can be retrieved and STVG coincides with GR in Solar System predictions.\par
	
	Both TeVeS and STVG seem to be good candidates for a new gravity theory. They both propose vectorial manifestations of the gravitational field. However, TeVeS has been recently criticized \cite{mavromatos2009,seifert2007} under light bending and stability considerations\footnote{Actually, there is discussion about these statements. For instance, Lasky argued that new generalizations of TeVeS, with equivalent solutions, may solve instability problems (see \cite{lasky2009,lasky2010}).}. On these subjects, STVG successfully describes strong and weak gravitational lensing by the Bullet Cluster \cite{brownstein2007}, but stellar structure stability has not been studied yet. For this reason, we begin the construction of stellar models in STVG. \par
	
	Most of STVG applications are based on vacuum solutions. In this paper, we find extended matter sourced solutions. We represent spacetime with a static, spherically symmetric manifold, and matter sources with a spherically symmetric perfect fluid energy-momentum tensor. We solve STVG field equations and analytically derive the metric components. Then, from STVG conservation equation, we obtain the modified Tolman-Oppenheimer-Volkoff equation (TOV). These results are used to construct STVG neutron star models. We integrate numerically the modified TOV equation for different neutron star equations of state (EoS) and compare the outcomes with GR results. Numerical integration is carried out with a variable-step, fourth-order, Runge-Kutta method.\par
	
	This work is organized as follows: in Section \ref{s_action} we present STVG action and field equations, and explain certain simplifications valid for neutron stars stellar structure. Solutions for extended matter sources are given in Section \ref{s_solution}. Then, in Section \ref{s_TOV}, we derive the modified TOV equation for STVG. In Section \ref{s_EoS}, we present different neutron stars EoS and explain our numerical integration method. In Section \ref{s_results}, we show the resulting mass-radius relation, mass-central density relation, and pressure, density, and mass profiles for distinct STVG neutron stars models, with the corresponding discussions. Section \ref{s_conclusions} is devoted to our main conclusions.\par

\section{STVG action and field equations}
\label{s_action}
	The essential idea of Moffat's STVG theory can be understood from its weak field, static, spherically symmetric, and constant scalar fields approximation (see Ref. \cite{moffat2006}). In that regime, the radial acceleration of a test particle at a distance $r$ from a gravitational mass source $M$ results:
	\begin{equation}
	\label{eq_weakfield}
		a(r)=-\frac{G_\mathrm{N}(1+\alpha) M}{r^2} + \frac{G_{\mathrm{N}}\alpha M}{r^2} e^{-mr}  (1+mr),
	\end{equation}
	where we choose natural units, $G_{\mathrm{N}}$ denotes Newton gravitational constant, and $\alpha,m$ are free parameters of the theory. The first term of Eq. \ref{eq_weakfield} prevails at $r\rightarrow \infty$, and represents an enhanced gravitational attraction, quantified by $G=G_{\mathrm{N}}(1+\alpha)$. Such term served to explain galaxy rotation curves, light bending phenomena, and cosmological data, without dark matter. The second term is significant when $mr<<1$ and represents gravitational repulsion. This Yukawa-type force counteracts the enhanced attraction and, from the interplay, Newton gravitational constant arise at $mr<<1$ scales.\par
	
	Numerical values for $\alpha$ and $m$ depend on the central source mass $M$, exhibiting the scalar field nature of $G$ and $m$. To date, no functional solutions of such fields has been proposed. However, numerical values for a wide range of central masses have been determined by adjusting phenomenology. For a review of such values, see Fig. 2 of Ref. \cite{brownstein2007}, where $M_0=\alpha^2 M$, and $r_0\sim m^{-1}$; as an example, for galaxies $\alpha \sim 10$ and $r_0 \ \sim 1\mathrm{kpc}$ (see Ref. \cite{moffat2013}).\par

	STVG action reads\footnote{Compared to Moffat's original action at Ref. \cite{moffat2006}, we nullify the cosmological constant because its effects are negligible over stellar structure, we ignore the scalar field nature of $\omega$ and set $\omega=1/\sqrt{12}$, as suggested by Moffat \cite{moffat2013,moffat2009}, and we set the potential $W(\phi)=0$ as is usually stated (see Ref. \cite{moffat2006}). Also, we propose a slight modification: we change the sign of vector field action $S_{\phi}$ in order to find agreement with the analogous Einstein-Maxwell formalism.}:
	\begin{equation}
	\label{totalaction}	
	S=S_{\mathrm{GR}}+S_{\phi}+S_{\mathrm{S}}+S_{\mathrm{M}},
	\end{equation}
where
	\begin{align}
S_{\mathrm{GR}}&= \frac{1}{16 \pi } \int d^4x \sqrt{-g} \frac{1}{G} R, \\
S_{\phi}&=  \omega \int d^4x \sqrt{-g} \left( \frac{1}{4} B^{\mu \nu} B_{\mu \nu} - \frac{1}{2} m^2 \phi^\mu \phi_\mu \right), \\
  &\begin{aligned}
   \hspace{-.57cm} S_{\mathrm{S}}&=  \int d^4x \sqrt{-g} \left[ \frac{1}{G^3} \left(\frac{1}{2} g^{\mu \nu} \nabla_\mu G \nabla_\nu G - V(G) \right) \right.+\\
      & \left. \frac{1}{G m^2} \left(\frac{1}{2} g^{\mu \nu} \nabla_\mu m \nabla_\nu m - V(m) \right)\right].
  \end{aligned}
\end{align}
Here, $g_{\mu \nu}$ denotes the spacetime metric, $R$ is the Ricci scalar, and $\nabla_{\mu}$ the covariant derivative; $\omega=1/\sqrt{12}$, $\phi^{\mu}$ denotes a Proca-type massive vector field, $m$ is its mass, and $B_{\mu \nu}=\partial_{\mu} \phi_{\nu} - \partial_{\nu} \phi_{\mu}$; $V(G),V(m)$ denote possible potentials for the scalar fields $G(x), \ m(x)$, respectively. We adopt the metric signature $\eta_{\mu \nu}=\mathrm{diag(}1,-1,-1,-1\mathrm{)}$, and choose natural units. The term $S_{\mathrm{M}}$ refers to possible matter sources.\par

	Since we are interested in the structure of compact neutron stars where $mr<<1$, we neglect the effects of the vector field mass $m$, and set $m=0$. From a physical point of view, we are not considering the decay of the Yukawa-type force because it happens far away from the gravitational source. This very same approximation is implemented, for instance, in Ref. \cite{moffat2015}. However, we remark that additional work is needed on this point in order to provide a fully satisfactory explanation of the behavior of STVG theory in the limit $m \rightarrow 0$, since discontinuities might appear in the corresponding solutions.
	
	There is too much freedom for the functional form of the scalar field $G$. In this paper, we focus on the vector field contributions to the field equations, and leave the study of different scalar field solutions of $G$ for a future work. Then, we adopt for the enhanced gravitational coupling constant $G$ the same prescription as Moffat \cite{moffat2006}:
\begin{equation}
\label{G}
G=G_{\mathrm{N}}(1+\alpha),
\end{equation}
and we sample the theory for different values of $\alpha$.\par

	Taking previous simplifications into account, the action \eqref{totalaction} reads:
\begin{equation}
\label{simpleaction}
	S=\int d^4x \sqrt{-g}\left[ \frac{1}{16 \pi G}  R +   \frac{\omega}{4} B^{\mu \nu} B_{\mu \nu} \right] + S_{\mathrm{M}}.
\end{equation}
Varying the latter with respect to the metric $g^{\mu \nu}$ yields:
\begin{equation}
\label{metriceq}
G_{\mu \nu} =  8\pi G \left( T^{\mathrm{M}}_{\mu \nu} + T^{\phi}_{\mu \nu} \right),
\end{equation}
where $G_{\mu \nu}$ denotes the Einstein tensor, and 
\begin{align}
\label{energymomentum1}
&T^{\mathrm{M}}_{\mu \nu} = -\frac{2}{\sqrt{-g}} \frac{\delta S_{\mathrm{M}}}{\delta g^{\mu \nu}}, \\
	&\begin{aligned}
	\label{energymomentum2}
	T^{\mathrm{\phi}}_{\mu \nu} &= -\frac{2}{\sqrt{-g}} \frac{\delta S_{\phi}}{\delta g^{\mu \nu}}= \\
	&=-\omega \left({B_\mu}^{\alpha} B_{\nu \alpha} - g_{\mu \nu} \frac{1}{4} B^{\rho \sigma} B_{\rho \sigma}		\right).
	\end{aligned}
\end{align}

Furthermore, varying the action \eqref{simpleaction} with respect to the vector field $\phi_{\mu}$ yields:
\begin{equation}
\label{vectoreq}
\nabla_{\nu} B^{\nu \mu} = \frac{1}{\omega} J^{\mu}_{\mathrm{Q}},
\end{equation}
where 
\begin{equation}
	J^{\mu}_{\mathrm{Q}}=-\frac{1}{\sqrt{-g}}\frac{\delta S_{\mathrm{M}}}{\delta \phi_\mu}=\sqrt{\alpha G_{\mathrm{N}}} J^{\mu}_{\mathrm{M}},
\end{equation}
i.e. the source of the vector field is the four-current matter density $J^\mu_{\mathrm{M}}$ multiplied by $\sqrt{\alpha G_{\mathrm{N}}}$. The latter constant is chosen to adjust observations (see, for instance, Eq. (13) of Ref. \cite{moffat2015}). The explicit functional dependence of $S_{\mathrm{M}}$ with $\phi_{\mu}$ has not been worked out so far. This is an important issue in the foundations of the theory that deserves further study. However, this point has no significant import for the application studied here and is left for a future publication.\par

	With the assumed approximations, Eqs. \eqref{metriceq} and \eqref{vectoreq} resemble Einstein-Maxwell equations. Their differences reside on the nature of the sources for the vector fields. In Einstein-Maxwell theory, mass and electric charge currents are independent properties of matter, and only the latter couples to the vector field $A^{\mu}$. In STVG, every massive current serves a source and couples to the vector field $\phi^{\mu}$. Then, a given matter distribution $T^{\mathrm{M}}_{\mu \nu}$ determines both the dynamics of the metric and vector fields.

\section{STVG static, spherically symmetric, matter sourced solution}
\label{s_solution}

	We model spacetime with a static, spherically symmetric geometry:
\begin{equation}
\label{metric}
	ds^2=e^{\nu(r)} dt^2 - e^{\lambda(r)} d r^2 - r^2 \left(d\theta^2+\sin^2\theta d\phi^2\right).
\end{equation}

	Further, we model the stellar matter content with a static, spherically symmetric, perfect fluid, energy-momentum tensor:
\begin{equation}
\label{energymomentummatter}
	{T^{\mathrm{M}}}_{\mu\nu} = \left[p(r)+\rho(r)\right] u_\mu u_\nu-p(r) g_{\mu\nu},
\end{equation}
where $p(r)$ and $\rho(r)$ denote the pressure and density of the fluid $r$-shell, respectively; $u^\mu\rightarrow \left(e^{-\nu/2}, 0, 0, 0\right)$ denotes the four-velocity of a mass element with $r$ coordinate. The corresponding four-current matter density is:
\begin{equation}
\label{mattercurrent}
	J^{\mu}_{\mathrm{M}}=4 \pi \rho u^\mu = \frac{ 4 \pi\rho}{\sqrt{g_{00}}} \frac{dx^{\mu}}{dx^{0}} \longrightarrow \left( 4 \pi\rho e^{-\nu/2} , 0, 0, 0\right).
\end{equation}

	Replacing the previous components of $J^\mu_{\mathrm{M}}$ in the vector field equation \eqref{vectoreq}, we obtain the non-vanishing components for $B^{\mu \nu}$:
\begin{equation}
	B^{41}=-B^{14}=-\frac{1}{\omega} \exp\left({-\frac{\nu+\lambda}{2}}\right) \frac{Q(r)}{r^2},
\end{equation}
where
\begin{equation}
	Q(r)\equiv \int dr e^{\lambda/2} \sqrt{\alpha G_{\mathrm{N}}} \rho 4 \pi  r^2 .
\end{equation}

	Then, non-vanishing components of the vector field energy-momentum tensor are:
\begin{equation}
\label{energymomentumvector}
	{{{T_\phi}^0}_0}={{{T_\phi}^1}_1}=-{{{T_\phi}^2}_2}=-{{{T_\phi}^3}_3}=\frac{1}{2\omega} \frac{Q^2(r)}{r^4}.
\end{equation}

	Replacing the components of both energy-momentum tensors \eqref{energymomentummatter} and \eqref{energymomentumvector} in the field equations \eqref{metriceq} we obtain the differential equation system:
\begin{align}
\left[-\frac{\nu''}{2} + \frac{\lambda' \nu'}{4} - \frac{\nu'^2}{4}-\frac{\nu'-\lambda'}{2r}\right] e^{-\lambda}&=-\kappa p - \frac{\kappa}{2\omega} \frac{Q^2}{r^4} \\
\frac{\lambda'}{r} e^{-\lambda} + \frac{(1-e^{-\lambda})}{r^2}&=\kappa\rho + \frac{\kappa}{2\omega} \frac{Q^2}{r^4}, \\
-\frac{\nu'}{r}e^{-\lambda} + \frac{(1-e^{-\lambda})}{r^2}&=-\kappa p + \frac{\kappa}{2\omega} \frac{Q^2}{r^4}.
\end{align}
Solutions for the latter system are given by:
\begin{equation}
\label{lambda}
e^{-\lambda(r)}=1-\frac{2G M(r)}{c^2 r} - \frac{1}{r} \frac{4 \pi G}{c^4\omega} \int dr \frac{Q^2(r)}{r^2},
\end{equation}
\begin{equation}
\label{nu}
\nu(r)=-\lambda(r)+\frac{8 \pi G}{c^4} \int dr e^{\lambda(r)} r \left(c^2 \rho(r) + p(r)\right),
\end{equation}
where we have recovered the speed of light factors. Notice that, for a point mass source, we retrieve Moffat's spherically symmetric black hole solution \cite{moffat2015}. The deduction presented in this section is similar to the work of P. Florides \cite{florides1983} for Einstein-Maxwell equations.\par

\section{Modified Tolman-Oppenheimer-Volkoff equation}
\label{s_TOV}

	STVG conservation equations reads:
\begin{equation}
\nabla_\mu \left( T_{\mathrm{M}}^{\mu \nu} + T_{\phi}^{\mu \nu} \right) = 0.
\end{equation}
Replacing the Christoffel symbols of metric \eqref{metric} in the $r$-component of the previous covariant derivative yields:
\begin{equation}
\label{TOV0}
\frac{dP(r)}{dr} + \frac{d}{dr} \left(-\frac{Q^2(r)}{2\omega r^4} \right)+\frac{\nu'(r)}{2} \left(p(r)+\rho(r)\right)-\frac{2 Q^2(r)}{\omega r^5}=0.
\end{equation}
From Eqs. \eqref{lambda} and \eqref{nu} we obtain an expression from $\nu'(r)$, and substitute it in \eqref{TOV0}. Isolating the radial pressure derivative we obtain the modified TOV equation:
\begin{equation}
\label{TOV}
\begin{aligned}
\frac{dP(r)}{dr} =& - \frac{e^{\lambda(r)}}{r^2} \left( \frac{4\pi G}{c^4} p(r) r^3 - \frac{2GQ^2(r)}{\omega c^4 r} + \frac{G M(r)}{c^2} +\frac{2 \pi G}{\omega c^4} \int dr \frac{Q^2(r)}{r^2} \right) \times  \\
&\times \left(\rho(r) c^2 + p(r) \right) +\frac{Q(r)}{w r^4} \frac{dQ(r)}{dr},
\end{aligned}
\end{equation}
where we have retrieved the speed of light factors. Because of the similarity of STVG and Einstein-Maxwell field equations, Eq. \eqref{TOV} can be compared with results of Ref. \cite{maurya2015} for electromagnetic mass models.\par

	While the enhanced gravitational constant $G$ causes a steepening of the pressure profile, $Q$-terms have opposite sign and tend to slow down the decrease. These latter terms manifest the repulsive behavior of gravity in STVG.  Setting the parameter $\alpha=0$ nullifies every $Q$-term, $G$ reduces to $G_{\mathrm{N}}$, and the classic relativistic TOV equation is recovered.\par	

\section{Equations of state and numerical integration}
\label{s_EoS}

	In order to apply the modified TOV equation to neutron star structure, we need an appropriate equation of state (EoS) that relates the pressure $P(r)$ with the density $\rho(r)$ of each $r$-shell. We consider four distinct neutron stars EoS denoted: POLY \cite{silbar2004}, SLY \cite{douchin2001}, FPS \cite{pandharipande1989} and BSK21 \cite{goriely2010,pearson2011,pearson2012}.\par

	POLY serves us to construct neutron star toy models. The EoS reads:
\begin{equation}
\label{POLY}
\zeta=2\xi+5.29355,
\end{equation}
where $\xi=\log\left(\rho \left[\mathrm{g \ cm^{-3}}\right]\right)$, and  $\zeta=\log\left(P\left[\mathrm{dyn \ cm^{-2}}\right]\right)$.This EoS is mathematically simple an well-behaved. Any peculiarity that would result from this toy model would be because of the effects of STVG.\par

	On the other hand, SLy, FPS and BSK21 are realistic neutron star EoS. These equations are determined by precise nuclear strong interaction models, and are usually given in the form of tables. In order to avoid interpolation ambiguities, we make use of analytical representations derived by Haensel and Potekhin \cite{haensel2004}, in the cases of SLy and FPS, and by Potekhin \cite{potekhin2013} for BSK21.\par

	Having the neutron star EoS, we proceed to integrate Eq. \eqref{TOV} numerically applying a fourth-order-Runge-Kutta method \cite{press1992}. Integration is carried out from the neutron star center up to its surface, defined as the $r$-shell where $\log\left( \rho_{\mathrm{s}}\left[\mathrm{g \ cm^{-3}}\right]\right)=6$. We assume central densities in the range: $14.6<\log\left(\rho_{\mathrm{c}}\left[\mathrm{g \ cm^{-3}}\right]\right)<15.9$. The integration method is taken from Ref. \cite{orellana2013}.\par

	Deviations from GR are expected for non-vanishing $\alpha$ (see Eq. \eqref{G}). This parameter mediates both, gravitational repulsion and enhanced attraction. In order to find agreement of STVG predictions with the perihelion advance of Mercury, Moffat determined for solar massive sources the upper limit \cite{moffat2006}:
\begin{equation}
\label{Sunupperlimit}
\alpha_\odot << \frac{1.5 \times 10^{5} c^2}{G_{\mathrm{N}}} \frac{1}{M_\odot} \mathrm{cm}.
\end{equation}

	Neutron stars have a few solar masses. Hence, we explore the restriction given by the inequality \eqref{Sunupperlimit} with neutron star structure. Formally, inside the star, $\alpha_{\mathrm{NS}}$ would depend on the contained mass of each $r$-shell. We take into account this dependence with a linear ad hoc prescription. Besides, we define a normalized factor $\gamma \in [0;1)$ to sample different values of $\alpha$:
\begin{equation}
\label{NSupperlimit}
\alpha_{\mathrm{NS}} = \gamma \frac{1.5 \times 10^{5} c^2}{G_{\mathrm{N}}} \frac{1}{M_\odot} \left(\frac{M(r)}{M_\odot}\right) \mathrm{cm},
\end{equation}
where $M(r)$ is the mass of the neutron star up to the $r$-shell. From this definition, STVG coincides with GR if $\gamma=0$, and Moffat's upper limit given by inequality \eqref{NSupperlimit} corresponds to $\gamma=1$.\par

\section{Results and discussion}
\label{s_results}

	The first result of our work is analytical and involves Eq. \eqref{TOV}, the modified TOV equation for STVG. The result shows clearly the functioning of STVG gravity: enhanced gravitational attraction, represented by negative terms in the pressure derivative, is counteracted by gravitational repulsion, the latter represented by positive $Q$-terms. Recall that the contributions of the scalar fields to Eq. \eqref{TOV} have not been considered . Future work will be devoted to the formal insertion of such Brans-Dicke type of fields.\par

	Further results include STVG neutron stars model. We integrate Eq. \eqref{TOV} for each EoS, for distinct values of the normalized parameter: $\gamma=0$ (GR)$, \ \gamma=1\times10^{-3}$, $\gamma= 2\times10^{-3}$, and $\gamma=4\times 10^{-3}$. Greater values of $\gamma$ yield unusual results that we will discuss later, while smaller values do not present significant deviations from GR.\par

\begin{figure*}
	\centering
  \includegraphics[width=\hsize]{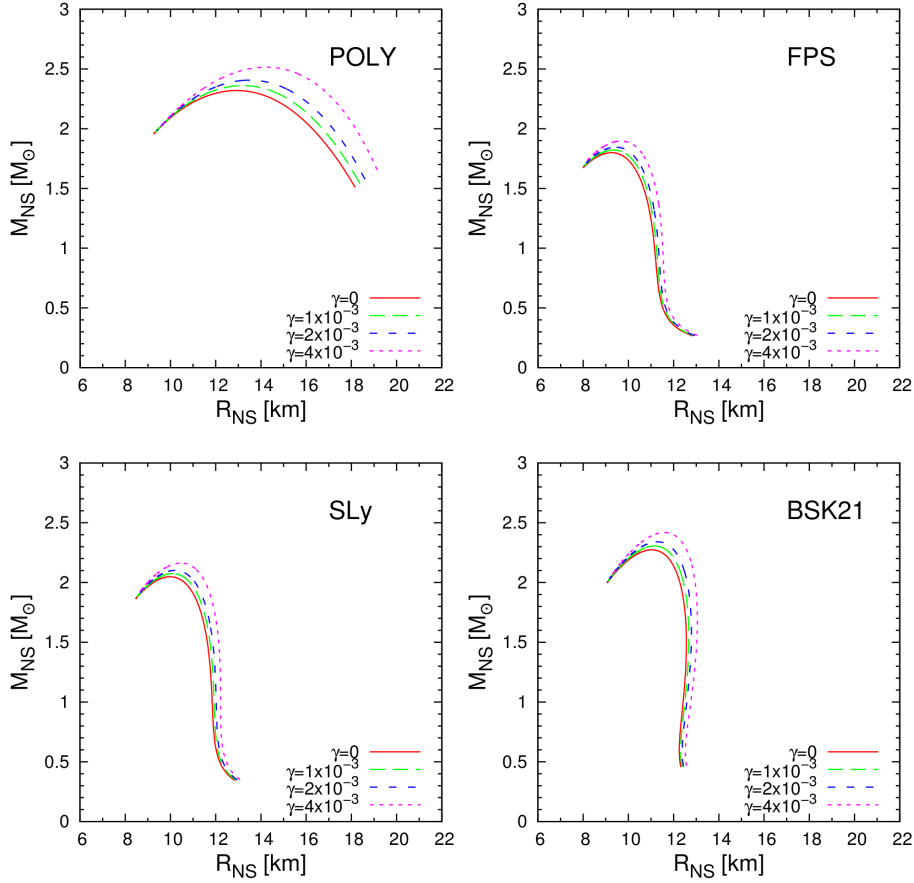}
	\vspace{.2cm}
  \caption{Mass-radius relations for neutron star models with central densities in the range: $14.6<\log\left(\rho_{\mathrm{c}}\left[\mathrm{g \ cm^{-3}}\right]\right)<15.9$. We consider four different EoS: POLY (\textit{top, left}), FPS (\textit{top, right}), Sly (\textit{bottom, left}) and BSK21 (\textit{bottom, right}). The parameter $\gamma$ quantifies the deviation of STVG from GR. In particular, $\gamma=0$ corresponds to GR predictions. As we can see, greater $\gamma$ implies higher maxima for neutron star total masses.}
	\label{fig:1}
\end{figure*}

\begin{figure*}
	\centering
  \includegraphics[width=\hsize]{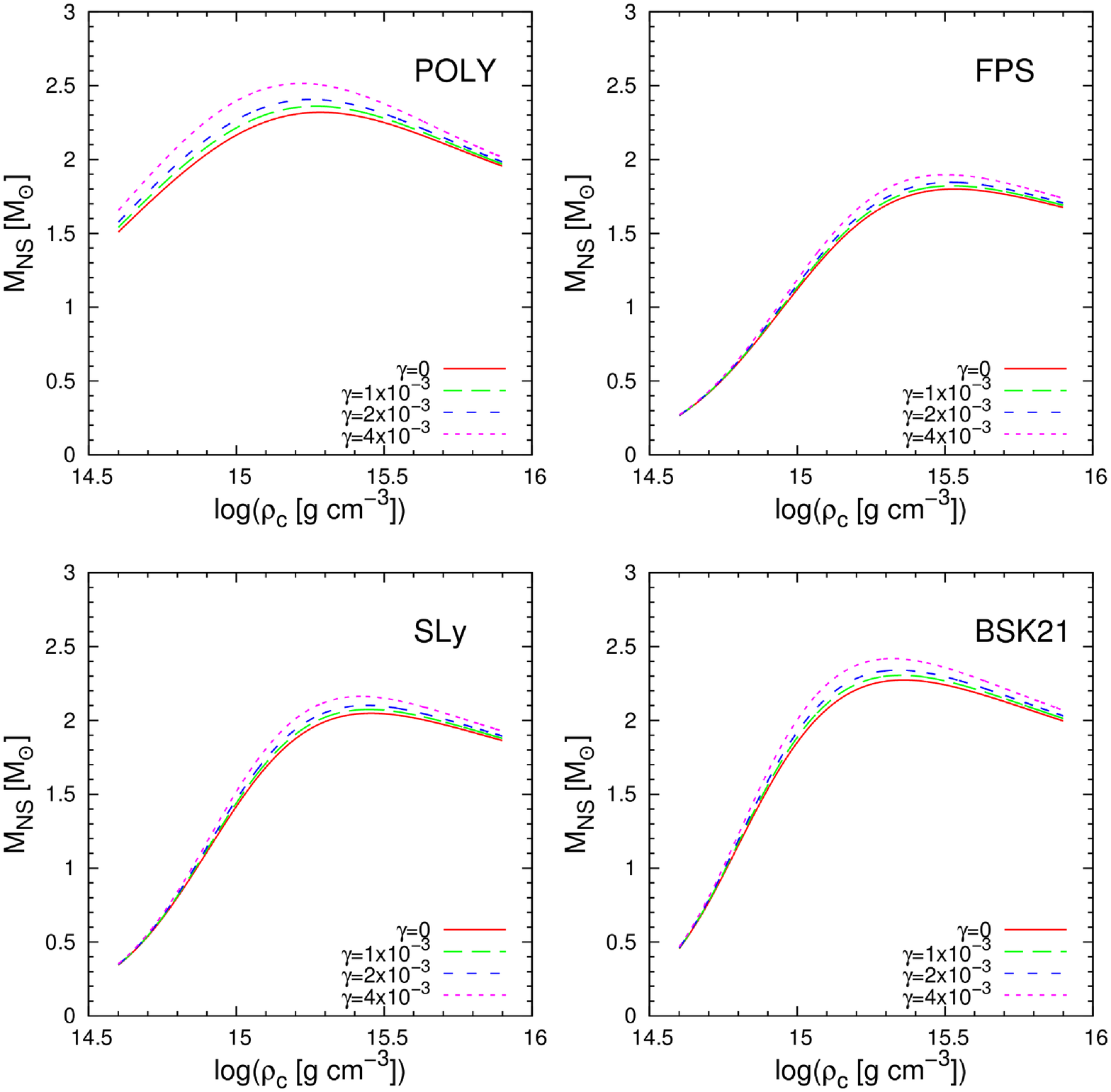}
	\vspace{.2cm}
  \caption{Relation between neutron star total masses and their central densities. As mentioned, the parameter $\gamma$ quantifies the deviation of STVG from GR. We consider four different EoS: POLY (\textit{top, left}), FPS (\textit{top, right}), Sly (\textit{bottom, left}) and BSK21 (\textit{bottom, right}). We find that the highest masses for STVG are achieved for lower central densities than in GR.}
	\label{fig:2}
\end{figure*}

	In Figure \ref{fig:1} we show the total mass-radius relation ($M_{\mathrm{NS}}-R_{\mathrm{NS}}$) for different central densities, for each EoS and $\gamma$ values. In Figure \ref{fig:2} we plot the total mass as a function of the central density. Relativistic results are in accordance with previous works like Ref. \cite{orellana2013}.\par

	From both Figs. \ref{fig:1} and \ref{fig:2} we notice that STVG coincides with GR predictions for high density neutron stars. Moffat's idea of retrieving classical results from the interchange of enhanced attraction and repulsion works properly in this context. However, differences arise as the central density decreases and finds its maximum at $\rho_{\mathrm{c}}\approx \times10^{15.3} \mathrm{g \ cm^{-3}}$. Remarkably, the curves tend to converge again for lower densities.\par

	We find that STVG neutron star maximal masses exceed GR results. Such STVG maxima are obtained for lower central densities than in GR, but the resulting star have larger radii. Recent astronomical determinations of neutron star masses \cite{antoniadis2013,kiziltan2013,ozel2012,demorest2012} defy relativistic limits. Hence, STVG stands as a strong candidate for a new gravity theory, at least in this aspect .\par

	STVG predictions are different from those of other alternative gravity theories. For instance in Ref. \cite{yazadjiev2014}, non-perturbative, and self consistent models of neutron stars have been constructed in squared-$f(R)$ gravity. For high mass neutron stars such models predict larger radii than GR, but for low mass neutron stars the behavior changes and final radii result smaller than the corresponding case in GR. In STVG, the first trend prevails along the whole mass range, i.e. final radii are always larger than the ones in GR. Contrary, as can be seen from Ref. \cite{lasky2008}, TeVeS models predict neutron stars with smaller masses and radii than GR, as a function of vector and scalar field coupling parameters.\par

\begin{figure*}
	\centering
	\includegraphics[width=\hsize]{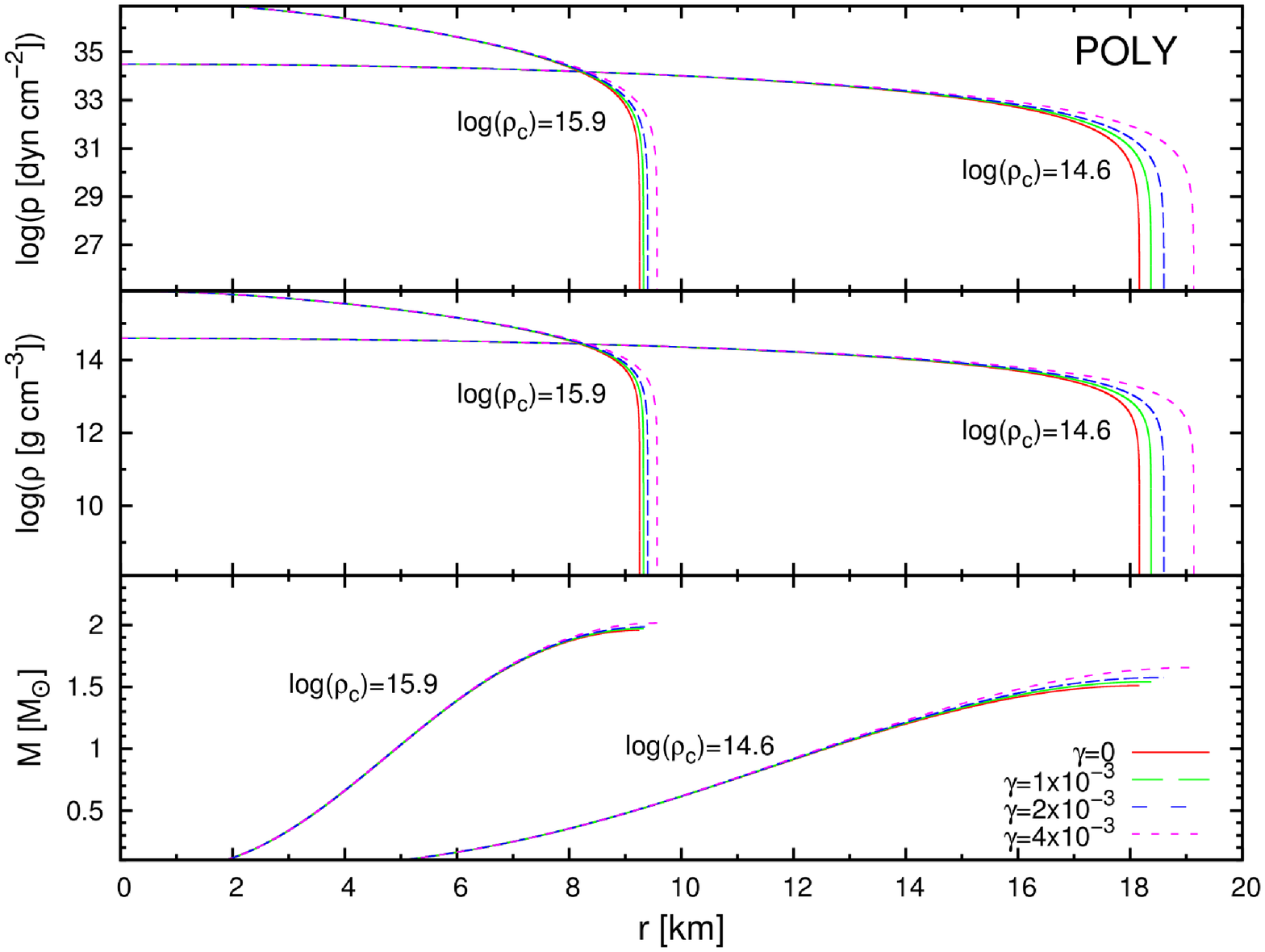}\vspace{.5cm}
	\includegraphics[width=\hsize]{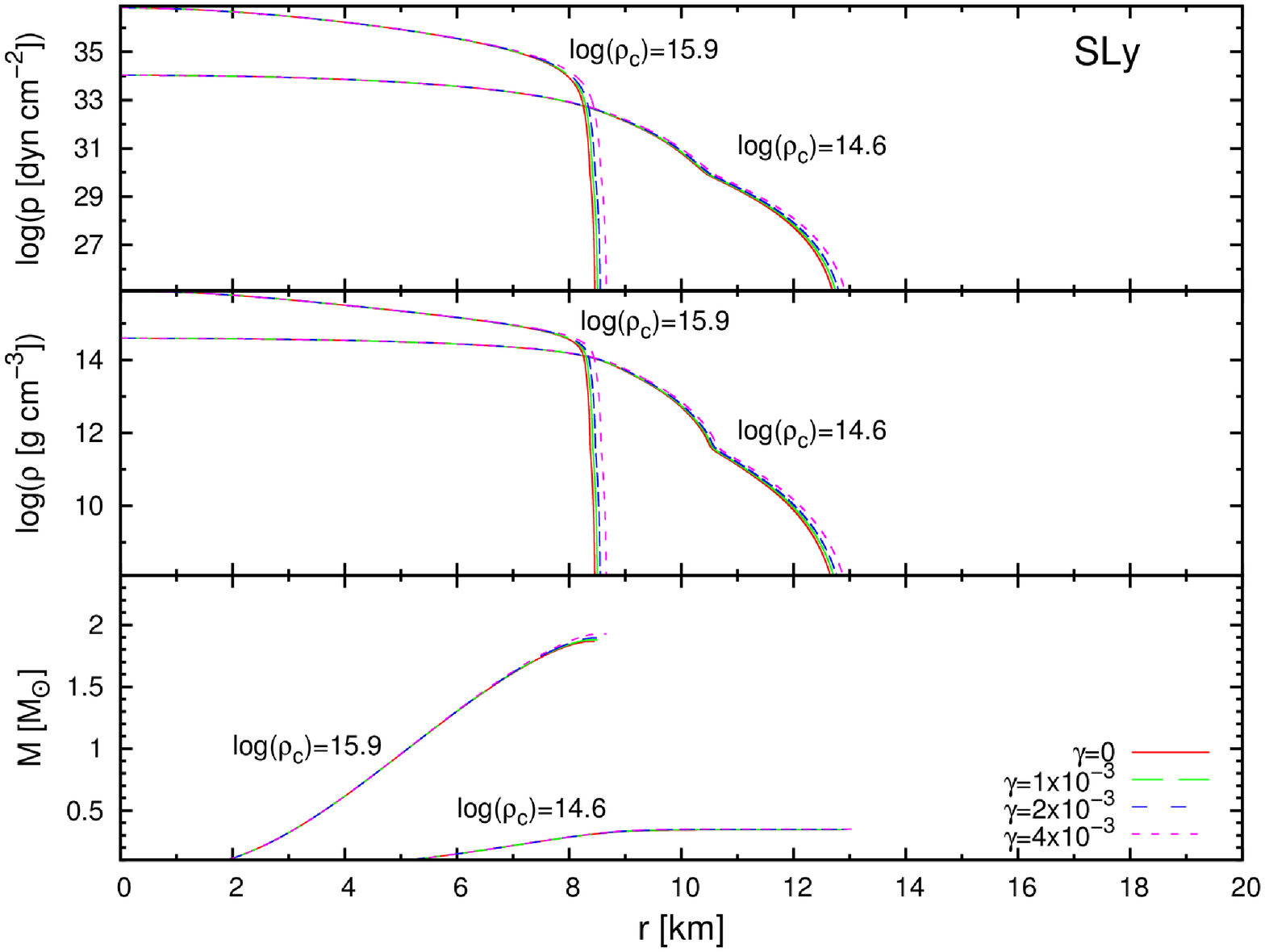}
	\vspace{.2cm}
  \caption{Pressure, density, and mass profiles for neutron star models in STVG. We display POLY (\textit{top}) and SLy (\textit{bottom}) outcomes. Both FPS and BSK21 models yield similar results. The graphics manifest the effects of STVG repulsive gravity, slowing down the pressure and density decrease. Consequently, STVG mass profiles reach higher values than in GR.}
  \label{fig:3}
\end{figure*}

	In Figure \ref{fig:3} we display pressure, density, and mass profiles for POLY and SLy neutron star models. From those graphics, we notice that STVG density and pressure profiles decrease slower than the relativistic ones. This implies that STVG mass profiles reach higher final masses. We associate this behavior with positive $Q$-terms in Eq. \eqref{TOV}. FPS and BSK21 profiles manifest equivalent deviations.\par

\begin{figure*}
	\centering
  \includegraphics[width=\hsize]{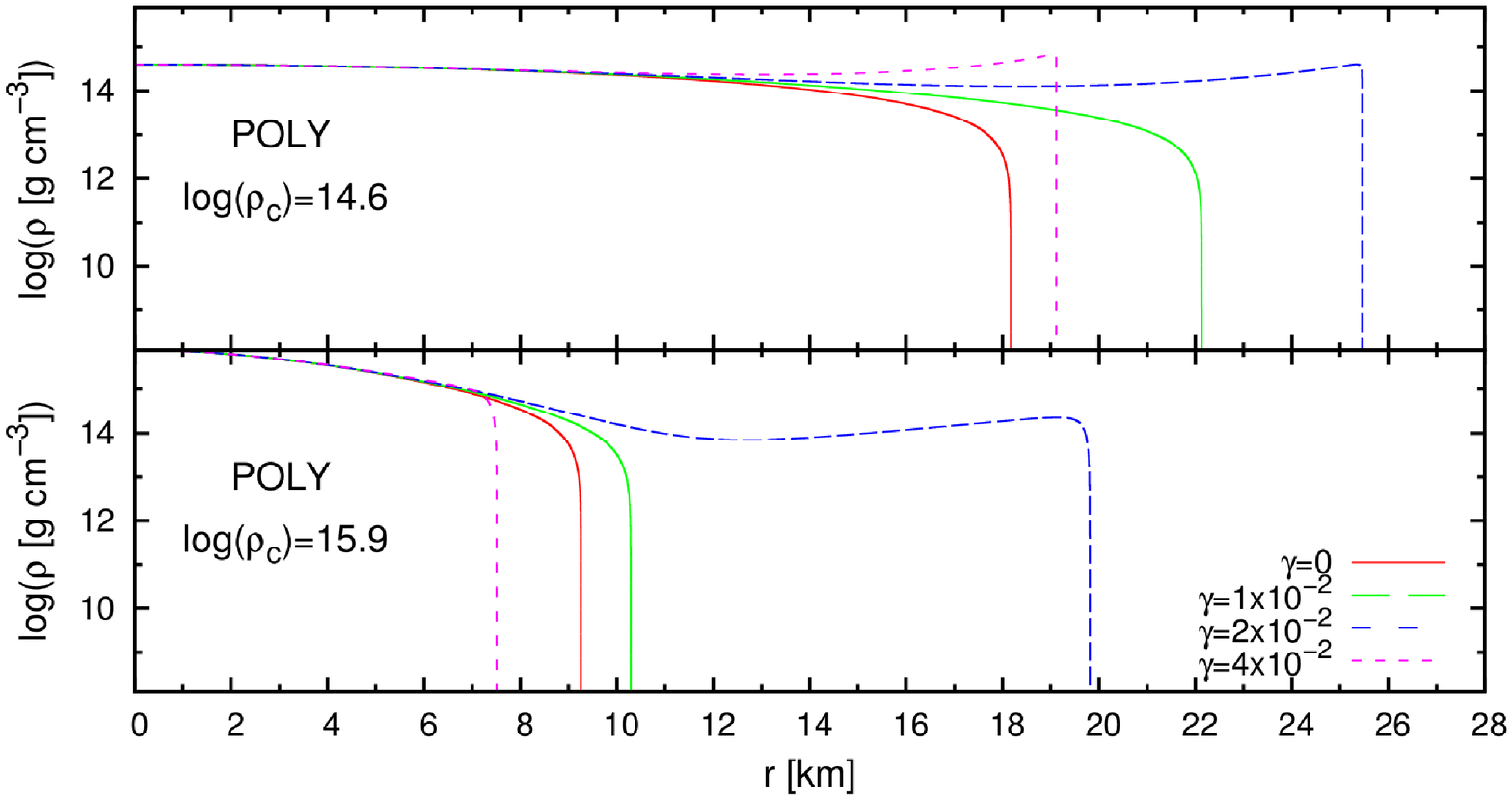}
	\vspace{.2cm}
  \caption{Density profiles for POLY neutron star models with central densities: $\log\left(\rho_{\mathrm{c}}\left[\mathrm{g \ cm^{-3}}\right]\right)=14.6$ (\textit{up}) and $\log\left(\rho_{\mathrm{c}}\left[\mathrm{g \ cm^{-3}}\right]\right)=15.9$ (\textit{down}). Significant differences between STVG and GR predictions arise from $\gamma \propto 10^{-2}$. Repulsive gravity effects are intensified in low density models, up to a certain limit, when enhanced attraction prevails. High density neutron star profiles show similar features but, enhanced attraction effects take place at smaller radii.}
	\label{fig:4}
\end{figure*}	

	Interesting results arise if we increase the parameter $\gamma$ by an order of magnitude. In Figure \ref{fig:4} we plot density profiles of POLY neutron stars with $\gamma=0, \ \gamma=1\times 10^{-2}, \ \gamma=2\times 10^{-2}$ and $\gamma=4\times 10^{-2}$. For low density neutron stars and great $\gamma$, repulsive gravity causes a peculiar growing profile. However, when a certain mass is reached, the enhanced attraction dominates and the profile decreases abruptly. A similar behavior is seen in high density neutron stars but, for the highest $\gamma$, enhanced attraction always prevails.\par

	Results for $\gamma\propto 10^{-2}$ suggest that, beyond the effects of repulsive gravity, for each $\gamma$ there is a limiting mass from where attraction dominates.  Peculiarities of these resulting density profiles may be useful when modeling pulsating stars, inhomogeneous cosmologies, or formation of galaxy filaments without dark matter.\par

	Within the approximations of our work, we find a tighter upper limit for the parameter $\alpha$. In order to obtain neutron stars with realistic masses, i.e. near astronomical determinations, and monotonically decreasing density profiles, we find that:
\begin{equation}
\label{newupperlimit}
\alpha < 10^{-2} \frac{1.5 \times 10^{5} c^2}{G_{\mathrm{N}}} \frac{1}{M_\odot} \mathrm{cm}.
\end{equation}
Recall that $\alpha$ responds to the dynamics of the scalar field $G$. The restriction given by inequality \eqref{newupperlimit} applies to the external field of solar-mass class of sources.

\section{Conclusions}
\label{s_conclusions}

	In this paper we presented a simplified version of Moffat's STVG field equations. We neglected the effects of the mass $m$ of the vector field because they manifest at large distances from the source, and approximated $G$ as a slowing varying constant. We warned the reader that the limit $m\rightarrow 0$ has not been theoretically explored, and may be discontinuous. Also, we commented that the explicit functional form of the matter action term has not been worked out so far. Formal work about these foundational issues of the theory is required and left for a separate research. We modeled a static and spherically symmetric matter distribution. Then, we solved the vector field equation and constructed the total energy-momentum tensor neglecting the contributions of scalar fields. After that, we solved the metric field equations, and derived the modified TOV equation from the conservation equation in STVG.\par

	We then constructed neutron star models. In order to integrate Eq. \eqref{TOV} we assumed four different EoS. We took the restriction for $\alpha_\odot$ parameter from Solar System observations and defined the normalized factor $\gamma$. Realistic deviations from GR arose from $\gamma\propto 10^{-3}$. The general feature of STVG pressure and density outcomes is a slower decreasing profile than in GR. This implies that STVG neutron star models admit heavier total masses than GR. The latter result is attractive because recent estimations of neutron star masses are defying GR limits.\par

	We incremented the parameter $\gamma$ by an order of magnitude. Distinctive local maxima and minima arose from the interchange between STVG gravitational repulsion and enhanced attraction. We consequently propose a more restrictive upper limit for the parameter $\alpha$ than the one obtained from Solar System observations.\par

	We conclude that, under certain simplifications, STVG entails admissible spherically symmetric and static solutions with matter sources. The corresponding modified TOV equation serves to construct realistic stellar models with higher total masses than GR. Density profiles with local minima and maxima are distinctive properties of STVG predictions.\par

	We expect novel predictions when taking rotation into account, like has already been done for TeVeS and squared-$f(R)$ gravity \cite{sotani2010,yazadjiev2015,staykov2014}. In STVG, the presence of Lorentz-like forces will manifest characteristic departures from other gravity theories. Future work will be dedicated to efforts in that direction. Furthermore, we will formally take into account the scalar fields contributions, and study stability and quasinormal modes of spherically symmetric solutions.\par

%
%




\end{document}